\documentclass[aps,prb,reprint]{revtex4-2}

\usepackage{amsmath,amsfonts,amssymb,graphicx,braket,bbold,mathtools,array}
\usepackage{float}
\usepackage{multirow}
\usepackage[]{units}
\usepackage{hyperref}
\hypersetup{
    colorlinks=true,
    linkcolor=blue,
    filecolor=blue,      
    urlcolor=blue,
    citecolor=blue
}

\usepackage[table]{xcolor}

\usepackage[T1]{fontenc}
\usepackage{dsfont}
\usepackage{tikz}
\usetikzlibrary{calc,patterns,decorations.pathmorphing,decorations.markings, shapes, positioning, shapes.geometric}

\newcommand*{\figref}[2][]{%
  \hyperref[{fig:#2}]{%
   \ref*{fig:#2}%
    \ifx\\#1\\%
    \else #1%
    \fi
  }%
}

\begin{document}
\title{A natural heavy-hole flopping mode qubit in germanium}
\author{Philipp M. Mutter}
\email{philipp.mutter@uni-konstanz.de}
\author{Guido Burkard}
\email{guido.burkard@uni-konstanz.de}
\affiliation{Department of Physics, University of Konstanz, D-78457 Konstanz, Germany}

\begin{abstract}
Flopping mode qubits in double quantum dots (DQDs) allow for coherent spin-photon hybridization and fast qubit gates when coupled to either an alternating external or a quantized cavity electric field. To achieve this, however, electronic systems rely on synthetic spin-orbit interaction (SOI) by means of a magnetic field gradient as a coupling mechanism. Here we theoretically show that this challenging experimental setup can be avoided in heavy-hole (HH) systems in germanium (Ge) by utilizing the sizeable cubic Rashba SOI. We argue that the resulting natural flopping mode qubit possesses highly tunable spin coupling strengths that allow for one- and two-qubit gate times in the nanosecond range when the system is designed to function in an optimal operation mode which we quantify. 
\end{abstract}

\maketitle
\section{Introduction}
The potential of implementing and manipulating physical qubits in semiconductor systems was first recognized in the late twentieth century~\cite{Loss1998, Kane1998, Imamoglu1999, Burkard1999, Kloeffel_review2013} and has since been demonstrated in countless experiments~\cite{Hanson2007review, Awschalom2013review, Zhang_review2019}. In particular, strongly confined HH states in nano-scale quantum systems have been investigated theoretically~\cite{Bulaev2005, Bulaev2005b, Bulaev2007, Kloeffel2011,  Kloeffel2013a, Kloeffel2013b, Kloeffel2018} and realized in recent years in various architectures ranging from one-dimensional hut- and nanowires to two-dimensional hole gases in planar heterostructures~\cite{Watzinger2018, Hendrickx2020, Hendrickx2020b, Hofmann2019arXiv, Wang2020arXiv, Froning2020arXiv, Lawrie2020arXiv, Jirovec2020arXiv}. Long coherence times due to the weak interaction with the host atomic nuclear spin bath and the absence of a valley degree of freedom in the valence band promote the binary pseudo-spin of such HH states to prime candidates for a reliable qubit~\cite{Scappucci2020arXiv}.

In this paper we propose a qubit built from the spin of the bonding HH state in a planar DQD. Compared to single quantum dot (QD) systems \cite{Mutter2020cavitycontrol}, such so-called flopping mode qubits possess a large dipole coupling to an applied alternating or quantized cavity electric field, allowing for coupling strengths beyond the decoherence rate~\cite{Mi2017, Stockklauser2017, Bruhat2018}. An alternative qubit-cavity construction uses multi-electron exchange-only qubits~\cite{Landig2018}. Among the most promising candidates for a platform for flopping-mode spin qubits are electrons in Silicon and Carbon DQDs, which have seen detailed studies regarding their performance and decoherence properties~\cite{Benito2017, Mi2018, Samkharadze2018, Cubaynes2019, Benito2019a, Benito2019b, Croot2020}. In these systems a magnetic field gradient perpendicular to the QDQ axis is applied to achieve a coupling between bonding and anti-bonding states of different spin. The magnetic field gradient enables the electron spins to distinguish between different positions in space and may therefore be seen as synthetic SOI. While coherent spin rotations via a cavity field could be achieved within this setup, experimental realizations rely on micro-magnets and it is yet unclear how such systems can be scaled to include a large number of qubits. On contrast, we show that HHs in Ge can form a natural flopping mode qubit, i.e., one that does not require synthetic SOI via a magnetic field gradient. As a result, it is expected to be realizable with far less effort in the laboratory and less prone to errors due to imperfections in dot engineering than its conduction band counterpart.

The feature that allows us to achieve a natural HH flopping mode qubit is the cubic Rashba SOI. This type of SOI is characteristic for valence band states and stems from the electric field produced by the atomic nuclei experienced by the HHs in a quantum well which introduces structural inversion asymmetry along the growth direction. Starting from the semi-microscopic Hamiltonian of a HH in a DQD subject to the cubic Rashba SOI and an out-of-plane magnetic field, we derive the ground state Hamiltonian describing the one particle states in the left and right dot, and obtain an explicit expression for the tunnel matrix element. Additionally, we find terms describing spin-flip tunneling and intra-dot spin-flips which are induced by the SOI and the effect of excited orbitals. These terms allow for spin couplings in the bonding state when the system is coupled to a classical alternating or a quantized cavity electric field. We investigate one- and two-qubit gate times and find operation times in the nanosecond range. The performance of the device depends crucially on system parameters such as the applied magnetic field, the inter-dot distance, the strength of the Rashba SOI and the dot detuning, and may thus be optimized via quantum engineering.

The remainder of this paper is structured as follows: In Sec.~\ref{sec:HH_states_in_a_DQD} we introduce the HH system and derive the DQD Hamiltonian in the orbital ground state. Building on these results, we take into account the effect of excited orbitals in a perturbative fashion in Sec.~\ref{sec:effect_higher_orbitals}. In Sec.~\ref{sec:bonding_anti_bonding} we change into the basis of bonding and anti-bonding states and investigate the coupling of these states via electric fields. We proceed to look at special parameter cases and derive effective spin-photon couplings allowing for spin rotations in the bonding state in Sec.~\ref{sec:spin_coupling}. Finally, Sec.~\ref{sec:conclusion} provides a conclusion and an outlook on possible future research concerning natural flopping mode qubits.

\section{Low-energy heavy-hole States} \label{sec:HH_states_in_a_DQD}

\begin{figure}
		\includegraphics[scale=0.5]{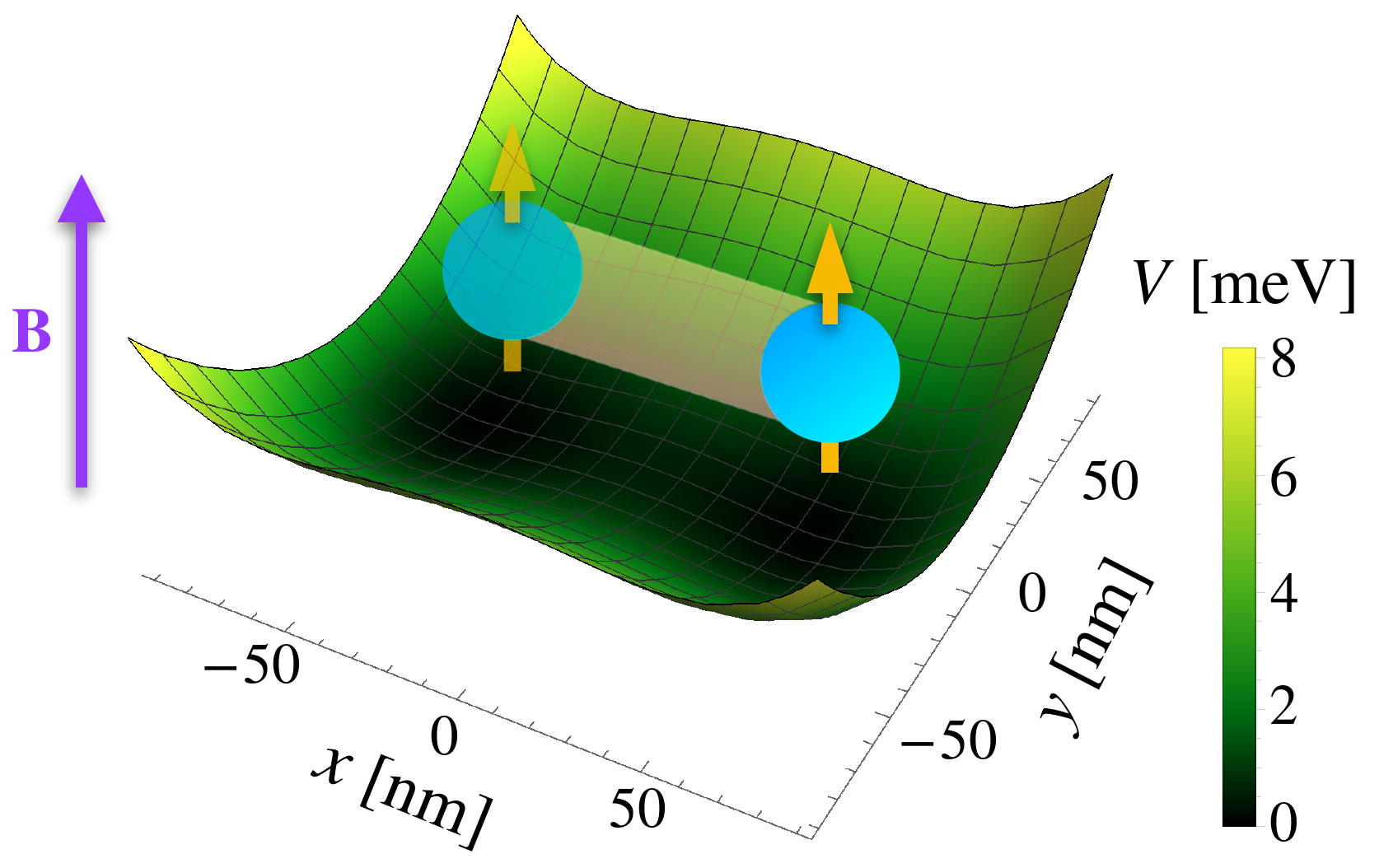}
		\caption{Potential landscape of a DQD as a function of the in-plane coordinates $x$ (along the DQD axis) and $y$ as given by Eq.~\eqref{eq:DQD_potential} with  $m = 0.1m_0$, $\hbar \omega_0 = 1$~meV and $a = 50$~nm. On top of the potential surface, we show a schematic of the flopping mode qubit, i.e., a single HH spin in the bonding state subject to an out-of-plane magnetic field $B$.}
		\label{fig:DQD_potential}
	\end{figure}

We consider a semiconductor heterostructure with a Ge quantum well such that the HHs in the material are subject to strong confinement along the out-of-plane ($z$) axis. To model an in-plane DQD we introduce a quartic, locally harmonic confining potential $V$ forming a double quantum well with dot separation $2a$ (Fig.~\ref{fig:DQD_potential}). Moreover, we allow for a static out-of-plane magnetic field $B$ and take into account the cubic Rashba SOI such that the system may be described by the Hamiltonian $H = H_0 + H_R$~\cite{Bulaev2005, Bulaev2005b, Bulaev2007},
	\begin{subequations}
	\begin{align}
	\label{eq:H_DQD}
		&H_0 = \frac{\pi_x^2 + \pi_y^2}{2m} + V(x,y) + \frac{g}{2} \mu_B B \sigma_z, \\
	\label{eq:DQD_potential}
		&V(x,y) = \frac{1}{2}m  \omega_0^2 \left( \frac{(x^2 - a^2)^2}{4a^2} + y^2 \right) , \\
	\label{eq:H_SO}
		&H_R = i \lambda_R \left( \sigma_+ \pi_-^3 - \sigma_- \pi_+^3 \right),
	\end{align}
	\end{subequations}
where $\boldsymbol{\pi} = \mathbf{p} + e  \mathbf{A}$ is the canonical momentum, $m$ the in-plane HH mass, $g >0$ the out-of-plane HH g-factor and $\sigma_z$ the Pauli matrix along the quantization axis. The cubic Rashba term $H_R$ features the spin ladder operators $\sigma_{\pm} = (\sigma_x \pm i \sigma_y)/2$ with in-plane Pauli matrices $\sigma_{x/y}$, $\pi_{\pm} = \pi_x \pm i \pi_y$ and the spin-orbit parameter $\lambda_R = 3 \gamma_s \alpha_R \langle E_z \rangle/2 m_0 \Xi$, where $\gamma_s = 5.11$ is the Luttinger parameter in spherical approximation, $\alpha_R$ the Rashba coefficient, $\langle E_z \rangle$ the average electric field induced by the structural inversion asymmetry due to out-of-plane confinement, $m_0$ the bare electron mass and $\Xi$ the HH-light hole splitting. There exists another  cubic Rashba term which turns out to be far less sizeable and may be neglected completely for the case of the valence band states in Ge where the spherical approximation applies~\cite{Marcellina2017}.
	
The time-independent Schr{\"o}dinger equation with the Hamiltonian~\eqref{eq:H_DQD} may be solved exactly close to the QD centres $x = \pm a$, where the quartic potential~\eqref{eq:DQD_potential} becomes locally harmonic. In locally symmetric gauge, $\mathbf{A}_{L/R} = B(-y,x \pm a,0)/2$, we obtain Fock-Darwin states $\psi_{nl}^{\text{FD}}$ (Appendix~\ref{appx:Fock-Darwin_states}) shifted to $x =  -a$ ($x = a$) for the left (right) dot with energies $E_{nl} = l \hbar \omega_L + (n+1) \hbar \omega $, where $\omega = \sqrt{\omega_0^2 + \omega_L^2}$ and $ \omega_L = eB/2m$ denotes the Larmor frequency. In what follows we work in the most general gauge for confinement and constant magnetic field along $z$,
	\begin{align}
	\label{eq:general_gauge}
		\mathbf{A} = \left(
			-B r y + \frac{\hbar}{e l_B }c_x,
			B(1-r)x + \frac{\hbar}{e l_B } c_y,
			0
		\right),
	\end{align}
where $r, c_x,c_y$ are arbitrary real constants, and we introduce the factor $\hbar /el_B$ including the magnetic length $l_B = \sqrt{\hbar/eB}$ such that $r, c_x,c_y$ are dimensionless. When transforming the left and right dot wave functions into this gauge, they acquire a magnetic phase and read
	\begin{align} 
	\begin{split}
		\psi_{nl}^{L/R} (x,y) = e^{  iy \frac{ (1-2r)x \pm a }{2l_B^2} } e^{ i \frac{c_xx+c_yy}{l_B}  } \psi_{nl}^{\text{FD}}(x \pm a,y).
	\end{split}
	\end{align}
Since there is a finite overlap
	\begin{align}
	\label{eq:overlap}
	S = \langle \psi_{00}^{L} \vert \psi_{00}^{R} \rangle = \exp \left( - \frac{a^2m(\omega_0^2 + 2 \omega_L^2)}{\hbar \omega}\right)
	\end{align}	
between the single QD ground states, we orthogonalize to obtain the Wannier states $ \vert L/R \rangle = \sqrt{N} \left(\psi_{00}^{L/R} - \gamma \psi_{00}^{R/L} \right)$, where the normalization factor is given by $N = \left(1-2 \gamma S + \gamma^2 \right)^{-1}$ and $ \gamma = \left(1- \sqrt{1-S^2} \right)/S$~\cite{Wannier1937, Wannier1962}. Defining the Wannier-spin states $\vert L/R, s \rangle \equiv \vert L/R \rangle \vert s \rangle$ with pseudo-spin quantum number $s \in \lbrace \uparrow, \downarrow \rbrace$ and the Pauli operators for position ($\tau$) and spin ($\sigma$), we obtain for the HH Hamiltonian $H$ in the orbital ground state basis $ \lbrace \vert L,\uparrow \rangle, \vert L,\downarrow \rangle, \vert R,\uparrow \rangle, \vert R,\downarrow \rangle \rbrace$,
	\begin{align}
		\label{eq:H_DQD_LR}
		H = -\frac{\epsilon}{2} \tau_z - t_c \tau_x + \frac{g}{2}   \mu_B B \sigma_z + \tilde{\eta} \tau_y  \sigma_y, 
	\end{align}
where
	\begin{align}
	\label{eq:matrix_elements}
	\begin{split}
		& t_c =  \frac{3N\gamma}{4} \left( \frac{ \hbar \omega_0^2}{ \omega }  +a^2m  \omega_0^2 \right),\\
		&\tilde{\eta}  =\lambda_R  N  (1 - \gamma^2)S   \left( \frac{am\omega_0^2 }{\omega} \right)^3.
	\end{split}
	\end{align}
The energy detuning between the two dots $\epsilon$ is applied via gate voltages in experiments and we add it as a phenomenological control parameter. On a more microscopic level there arises a spin-conserving tunnel matrix element $t_c$ due to the DQD potential. On the other hand, there is a spin-flip tunneling term $\tilde{\eta}$ due to the Rashba SOI. This term plays an important role in the ground state spin coupling in the bonding state discussed in the following.

As a consequence of the U(1) gauge invariance of quantum electrodynamics all observable quantities obtained are gauge-invariant, i.e., they do not depend on the constants $r$, $c_x$ and $c_y$. Additional constant gauge phases, which only affect the gauge field $\Lambda$ but do not change the vector potential under the transformation $\mathbf{A} \rightarrow \mathbf{A}  + \nabla \Lambda$, can be undone by choosing an appropriate representative from the ray of states in the Hilbert space for the wave function in each dot. Since the dependence of the tunnel coupling $t_c$ on the magnetic field exceeds a pure phase shift~\footnote{Indeed, the effect is even more pronounced than in comparable electronic systems due to the low in-plane HH mass, $m \simeq 0.1 m_0$, where $m_0$ is the bare electron mass.} (Fig.~\ref{fig:comparison_matrix_elements}), our results go beyond the standard interpretation of Peierls substitution prescription~\footnote{In his original work, Peierls himself pointed out that the effect of a non-zero magnetic field exceeds a pure phase factor~\cite{Peierls1933}.}, which always ought to be treated with the appropriate care~\cite{Alexandrov1991, Alexandrov1991b, Ibanez2014}. Indeed, the validity of the approximation stating that the tunnel elements are only changed by a phase factor upon the application of an external magnetic field relies on strongly localized Wannier functions such that the area $A_{\psi}$ they enclose in the $x$-$y$-plane is much smaller than the area of a flux quantum, $A_B = \hbar \pi /eB$. The DQD potential discussed in the present work yields local eigenfunctions that do not satisfy this assumption. To be specific, the mean area of the Fock-Darwin ground state $\psi_{00}^{\text{FD}}$ is estimated to be $A_{\psi} = \pi \langle \psi_{00}^{\text{FD}} \vert x^2 \vert \psi_{00}^{\text{FD}} \rangle \sim \pi \hbar / m \omega$, and thus $A_{\psi} / A_B \sim \omega_L/\omega \sim 0.1$ at $B = 1$~T.

\section{Effect of higher orbital states}
\label{sec:effect_higher_orbitals}

To include the effects of higher orbital states we consider the SOI as a small perturbation and work with the  single-dot ground states within first-order perturbation theory,
	\begin{align}
	\begin{split}
		&\vert \tilde{\psi}_{00\Uparrow} \rangle = \vert \psi_{00 \uparrow} \rangle + \frac{\lambda_R \sqrt{6} \omega_+^3 (\hbar m /\omega)^{3/2}}{3 \hbar \omega_+ - \hbar \omega_Z} \vert \psi_{33 \downarrow} \rangle, \\
		&\vert \tilde{\psi}_{00\Downarrow} \rangle = \vert \psi_{00 \downarrow} \rangle - \frac{\lambda_R \sqrt{6} \omega_-^3 (\hbar m /\omega)^{3/2}}{3 \hbar \omega_- + \hbar \omega_Z } \vert \psi_{3-3 \uparrow} \rangle,
	\end{split}
	\end{align}
where $\hbar \omega_Z = g \mu_B B$ and $\omega_{\pm} = \omega  \pm \omega_L$. Note that the normalization is omitted as it differs from unity only in second order in the small perturbation parameter $\xi_R = \lambda_R (\hbar m \omega_0)^{3/2}/(3\hbar \omega_- + \hbar \omega_Z) $. The perturbative approach is valid as long as $\vert \xi_R \vert \ll 1$, which holds for all relevant magnetic field strengths, $B \lesssim 100$~T for $\hbar\alpha_R \langle E_z \rangle = 10^{-11} $~eVm and $\Xi = 100$~meV. Due to the cubic SOI the altered ground states each contain a component with opposite spin in the third excited orbital level.  As in Sec.~\ref{sec:HH_states_in_a_DQD} the states in the left and right dot are obtained by shifting the position coordinate to the dot centres $x \rightarrow x \pm a$ and gauge transforming back to common gauge, Eq.~\eqref{eq:general_gauge},
\begin{align} 
	\begin{split}
		\tilde{\psi}^{L/R} (x,y) = e^{  iy \frac{ (1-2r)x \pm a }{2l_B^2} } e^{ i \frac{c_xx+c_yy}{l_B}  }  \tilde{\psi}_{00}(x \pm a, y).
	\end{split}
	\end{align}
We find the overlaps 
$S_{\Uparrow / \Downarrow}  = \langle  \tilde{\psi}^L_{\Uparrow / \Downarrow} \vert  \tilde{\psi}^R_{\Uparrow / \Downarrow} \rangle = S \left(1 + \mathcal{O}(\xi_R^2 ) \right)$,
where $\vert  \tilde{\psi}^{L,R}_{\Uparrow / \Downarrow} \rangle=\vert \tilde{\psi}^{L/R}\rangle\vert\Uparrow / \Downarrow\rangle$, and
	\begin{align}
	\begin{split}
		&S_{\Uparrow \Downarrow}  =  \left\langle  \tilde{\psi}^L_{\Uparrow} \vert  \tilde{\psi}^R_{\Downarrow} \right\rangle
		= S \frac{6 \lambda_R   m^3a^3 \omega_0^6}{\hbar \omega^2 \left( 9\omega_0^2+6 \omega_L \omega_Z - \omega_Z^2 \right)},
	\end{split}
	\end{align}
 $S_{\Downarrow \Uparrow}  =   - S_{\Uparrow \Downarrow}$, with $S$ as given in Eq.~\eqref{eq:overlap}. Neglecting the overlaps containing powers of $\xi_R$, we obtain the Wannier-spin states, which are orthogonal within our approximation,
	\begin{align}
	 \vert L/R, \sigma\rangle = \frac{ \big\vert \tilde{\psi}^{L/R}_{\sigma} \big\rangle - \gamma \big\vert \tilde{\psi}^{R/L}_{\sigma} \big\rangle}{\sqrt{1-2 \gamma S + \gamma^2}}.
	 \label{eq:Wannier_excited_orbitals}
	\end{align}
Here, the index $\sigma = \,\,\Uparrow,\Downarrow$ labels the constituents of a Kramers pair for each dot, which we refer to as a pseudo-spin doublet.

We proceed to project the HH DQD Hamiltonian $H = H_0 + H_R$ onto the Wannier basis $\lbrace {\vert L,\Uparrow \rangle}, {\vert L,\Downarrow \rangle}, {\vert R,\Uparrow \rangle}, {\vert R,\Downarrow \rangle \rbrace}$. To facilitate the integrals appearing in the calculations, we use the behaviour of the unperturbed left and right dot states including a magnetic phase in symmetric gauge under (partial) parity transformations,
	\begin{align}
	\begin{split}
		&\psi_{nl}^L( -x, -y) = (-1)^n \psi_{nl}^R(x,y), \\ 
		&\psi_{nl}^L(-x,y) = (-1)^n \psi_{nl}^R(x,y)^*.
	\end{split}\label{eq:transformationrules}
	\end{align}
The spin-conserving matrix elements and the out-of-plane Zeeman term receive corrections only at second-order in $\xi_R$ which we neglect. In contrast, non spin-conserving terms appear in first order in the SOI, and as a result the Hamiltonian features corrections to the spin-flip tunneling term $\tilde{\eta}$ as well as an extra intra-dot spin-flip term $\chi$,
	\begin{align}
	\begin{split}
	\label{eq:H_h}
		H = -\frac{\epsilon}{2} \tau_z - t_c \tau_x + \frac{g}{2}   \mu_B B \sigma_z + \eta  \tau_y  \sigma_y + \chi \tau_z \sigma_x,
	\end{split}
	\end{align}
where $  \eta =  \tilde{\eta} + \delta  \tilde{\eta}_1 + \delta  \tilde{\eta}_2+ \frac{g}{2} \mu_B b$, $\tilde{\eta} $, $ \epsilon$ and $t_c$ as in Eq.~\eqref{eq:matrix_elements}  and
	\begin{align}
	\label{eq:matrix_elements_excited_states}
	\begin{split}
		& \chi = \frac{\lambda_R \bar{N}  3m \omega_0^2 \hbar^2}{8 \omega^3 a}  \left( \frac{\omega_-^3}{3  \omega_- + \omega_Z } -  \frac{\omega_+^3}{3  \omega_+  -  \omega_Z } \right), \\
		&\delta \tilde{\eta}_1 = \lambda_R \bar{N} \sqrt{6} \left( \frac{\hbar m}{\omega} \right)^{\frac{3}{2}} \sum_{\pm} \frac{\omega_{\pm}^3\langle \psi_{00}^L \vert h_0 \vert \psi_{3 \pm 3}^R\rangle}{3 \hbar \omega_{\pm}   \mp \hbar \omega_Z },  \\
		&\delta \tilde{\eta}_2 =  \frac{6 \lambda_R S  m^3a^3 \omega_0^6}{ \omega^2 \left( 9\omega_0^2 + 6 \omega_L \omega_Z - \omega_Z^2 \right)} \left( \omega + \frac{3 \hbar \omega_0^2}{32 a^2 m\omega^2}  \right), \\
		& b = \lambda_R B \bar{N} S  \frac{a^3m^3 \omega_0^6}{\omega^3 \hbar}   \frac{2 \omega_Z - 6 \omega_L}{9 \omega_0^2 + 6  \omega_L \omega_Z-  \omega_Z^2}.
	\end{split}
	\end{align}
We make use of the abbreviations $\bar{N} = N(1-\gamma^2)$ and $h_0 = H_0 - g \mu_B B \sigma_z /2$, and the wavefunctions $\psi_{nl}^d$ are the unperturbed shifted Fock-Darwin states including a magnetic phase from transforming back to symmetric gauge. The matrix elements appearing in $\delta \tilde{\eta}_1$ are straightforward to calculate but lengthy; they are shown in Appendix~\ref{appx:spin_flip_matrix_elements}. Analysing the form of the novel terms, we find that the quantity $\chi$ acts as an intrinsic magnetic field gradient of strength $2 \chi$. Note that $\chi(-B)=-\chi(B)$ and hence $\chi(B=0)=0$. While the term arises in a perturbative approach in the SOI, its magnitude may be sizeable as it does not depend on the overlap $S$ between the left and right dot states. At $B=1$~T one has $2\chi\simeq 0.6$~$\mu$eV for a HH-light-hole splitting $\Xi = 100$~meV and a Rashba coefficient $ \hbar \alpha_R \langle E_z \rangle = 10^{-11} $~eVm, which is of the same order of magnitude as the magnetic field gradients applied in electronic systems, $\Delta B_x \sim 1.6$~$\mu$eV~\cite{Benito2017, Mi2018}. Finally, the corrections to the spin-flip tunneling matrix element $\tilde{\eta}$ are threefold: $\delta \tilde{\eta}_1$ arises due to the DQD potential and $b$ is due to the out-of-plane magnetic field in our approximation of zero overlap between different pseudo-spins. On the other hand,  $\delta \tilde{\eta}_2$ is due to a non-vanishing overlap between states of different pseudo-spin, and the only instance where our approximation breaks down. The $\delta \tilde{\eta}_2$ contribution can be treated within a different approximative approach as detailed in Appendix~\ref{appx:approximative_solution_Wannier}. A comparison of the matrix elements as a function of the magnetic field is shown in Fig.~\ref{fig:comparison_matrix_elements}. In the figure the analytical results are complemented by numerical data which takes into account all overlaps between the dot states and also second-order terms in the spin-orbit parameter $\xi_R$ (Appendix~\ref{appx:numerical_study}.). We find excellent agreement, a posteriori justifying the approximations made.

\section{Electric coupling of bonding and anti-bonding states} \label{sec:bonding_anti_bonding}

\subsection{Bonding-anti-bonding basis}

\begin{figure}
	\includegraphics[scale=0.23]{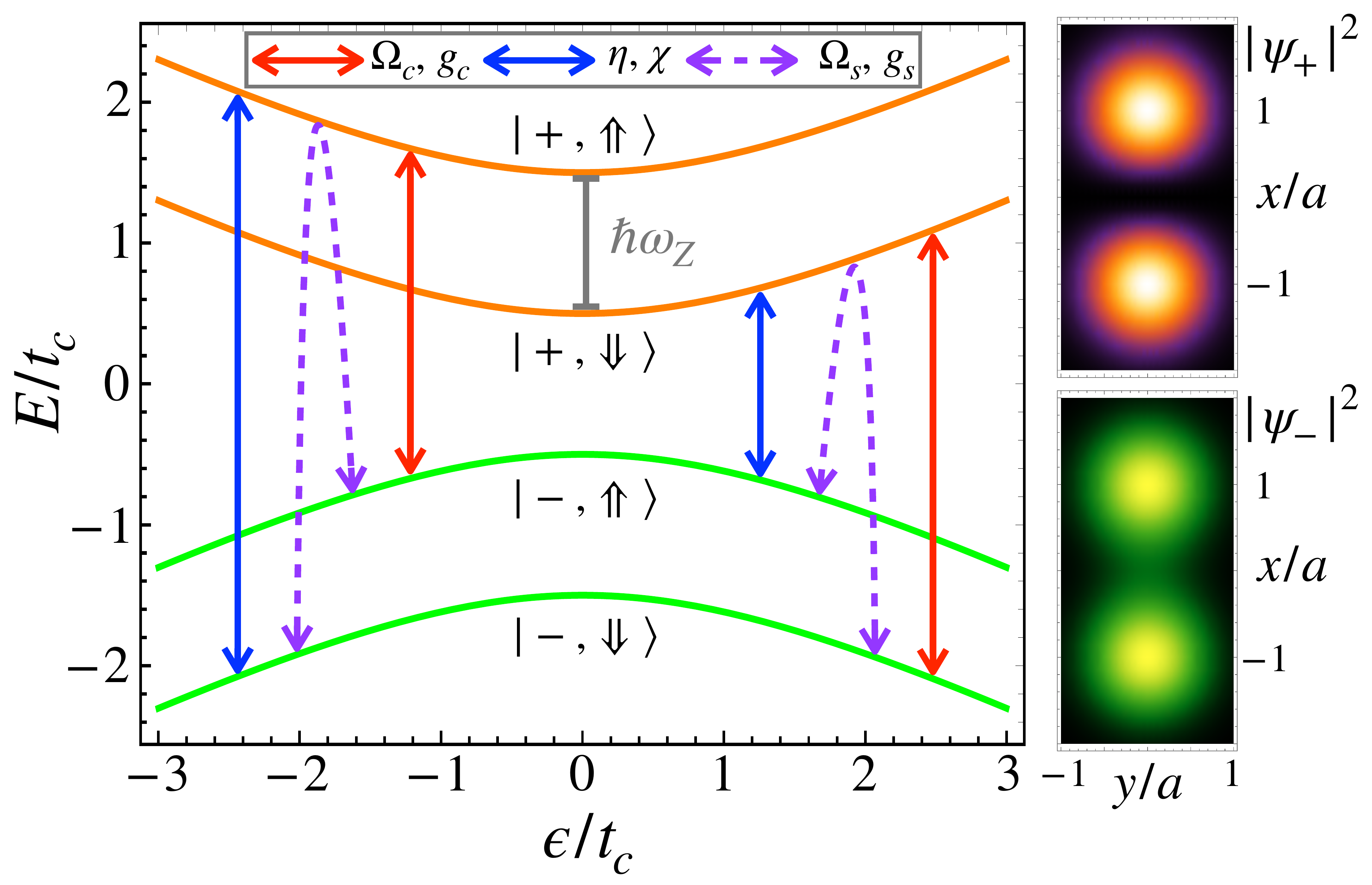}
	\caption{Spin coupling mechanism in the bonding state. We show the unperturbed Zeeman-split energies of the bonding (green lines) and anti-bonding (orange lines) states as a function of the detuning. The red and blue double-headed arrows represent transitions due to the dipole coupling and SOI, respectively, which can be combined to obtain an effective spin coupling in the bonding state (purple dashed arrows). On the right we display the probability densities of bonding (bottom) and anti-bonding (top) states according to the applied Wannier formalism at zero detuning. Brighter colors correspond to larger values.}
	\label{fig:bonding_anti_bonding_schematic}
	\end{figure}

It is useful to diagonalize the effective DQD Hamiltonian $H$, Eq.~\eqref{eq:H_h}, at zero magnetic field and vanishing SOI, yielding the eigenbasis $\lbrace \vert -,\Uparrow\rangle, \vert -,\Downarrow \rangle, \vert +,\Uparrow\rangle, \vert +,\Downarrow \rangle \rbrace$ with energies $E_{\pm} = \pm \sqrt{t_c^2 + \epsilon^2/4}$ (Appendix~\ref{appx:bonding_anti_bonding_basis}). Since the $\vert - \rangle$ states are lower in energy compared to the $\vert + \rangle$ states, we refer to them as bonding and anti-bonding, respectively, a nomenclature inspired by similar states in molecular physics. To understand the mixing of these states in the system under consideration, we proceed to transform the total HH Hamiltonian $H$ at non-zero magnetic field and in the presence of the SOI into the bonding-anti-bonding basis,
	\begin{align}
	\label{eq:H_GS_L_R}
		H = \begin{pmatrix}
			-K_- & \chi_s  & 0 &  - \eta -\chi_c \\
			\chi_s & -K_+ & \eta - \chi_c & 0 \\
			0 & \eta - \chi_c & K_+ & -\chi_s \\
			- \eta - \chi_c & 0 & -\chi_s& K_-
		\end{pmatrix},
	\end{align}
where we define $\chi_s = \chi \sin \theta$, $\chi_c = \chi \cos  \theta$ with the DQD orbital mixing angle $\theta =   \arctan \left( \epsilon/2t_c \right)$ and $K_{\pm} = \sqrt{t_c^2 + \epsilon^2/4} \pm  g \mu_B B /2$. As one can see, the SOI couples  orbital states of different pseudospin. Our aim is to complement this coupling with dipole transitions induced by an electric field, i.e., transitions between orbital states of equal spin, to obtain effective spin rotations in the orbital of lowest energy (Fig.~\ref{fig:bonding_anti_bonding_schematic}). 
\subsection{Coupling to electric fields}
Both  classical electric fields and quantum mechanical photons may be coupled to the DQD system. Neglecting diamagnetic contributions, the minimally coupled Hamiltonian reads
	\begin{align}
		H_I = \frac{e}{m} \mathbf{A}_{\text{e}} \cdot \boldsymbol{\pi}, 
	\end{align}
where the time-dependent vector potential $\mathbf{A}_{\text{e}}$ may describe a classical alternating electric field or photons within the framework of cavity quantum electrodynamics. In the present case the light must be polarized along the DQD axis $x$ to couple bonding and anti-bonding states, and we find for the dimensionless momentum operator $ \Pi_x =  \pi_x /\sqrt{m \hbar \omega_0}$,
	\begin{align}
\Pi_x =  d_c \tau_y + d_s \left( \frac{2\gamma}{1+\gamma^2} + \sin  \theta \tau_x  + \cos  \theta \tau_z   \right)  \sigma_y,
	\end{align}
where the operators for position ($\tau$) and spin ($\sigma$) act in the bonding-anti-bonding basis and the dipole elements read
	\begin{align}
	\label{eq:dimensionless_momentum}
	\begin{split}
			&d_c = N (1 - \gamma^2)S  \frac{ a \omega_0 }{r \omega}  + \mathcal{O} \left( \xi_R^2 \right), \\
			&d_s = \lambda_R  N (1+\gamma^2)  S \frac{m^2 a^2 \omega_0}{2 r\omega^2} \sum_{\pm} \pm \frac{\omega_{\pm}^3 C_{\mp}}{\omega_Z \mp 3 \omega_{\pm}}, \\
			& C_{\pm} = \frac{2a^2m}{\hbar \omega^3} \left( \omega_0^4 \pm 4 \omega_L^3 \omega_{\pm} +  \omega_0^2 \omega_L( 5 \omega_L \pm 3 \omega)  \right) - 3\frac{\omega_{\pm}}{\omega},
		\end{split}
	\end{align}
with the typical lateral QD size $r= \sqrt{ \hbar / m  \omega_0}$. Hence, the electromagnetic fields achieve a coupling $d_c$ between bonding and anti-bonding states of equal spin. Additionally, due to the effect of higher orbital states which are admixed due to the SOI, there are non spin-conserving transitions within a dot of strength $d_s$ and $d_s \cos  \theta$ as well as transitions between the two dots of strength $d_s \sin  \theta$.  For all cases the coupling strength is sensitive to the inter-dot distance and the applied magnetic field. We now turn to the description of the explicit coupling to a classical alternating and a quantized cavity electric field.

Electric dipole spin resonance (EDSR) aims to manipulate the spin degree of freedom with the aid of an alternating electric field and has been shown to be implementable in a variety of systems~\cite{Duckheim2006, Rancic2016, Brooks2020}. For the DQD system in this study, we assume a field of amplitude $\mathcal{E}$, driving frequency $\omega_d$ and polarization along the DQD axis $x$, $\boldsymbol{\mathcal{E}}(t) = -\mathcal{E} \cos (\omega_d t) \mathbf{e}_x$. The Hamiltonian describing this field is given by
	\begin{align}
	\label{eq:H_int_EDSR}
		H_{\mathcal{E}} = \frac{e\mathcal{E}}{ \omega_d} \sqrt{\frac{\hbar \omega_0}{m}} \sin (\omega_d t) \Pi_x \equiv \hbar \Omega_c \sin (\omega_d t) \Pi_x,
	\end{align}
where we define the coupling strength $\hbar \Omega_c = e\mathcal{E}r \omega_0/ \omega_d$ for later convenience.

Finally, we turn to the description of a quantized cavity field. A single cavity mode of frequency $\omega_c$ is described by the Hamiltonian $H_c = \hbar \omega_c b^{\dagger} b$, where $b^{\dagger} $ and $b$ denote the photon creation and annihilation operator, respectively. The interaction of the cavity photons with the HHs confined in the DQD for linearly polarized light along $x$ is of the form
	\begin{align}
	\label{eq:H_int_cavity}
	H_c = \hbar g_c (b + b^{\dagger}) \Pi_x,
	\end{align}
where $g_c = e\sqrt{\omega_0/2 \epsilon_0 \epsilon_r m V \omega_c}$ is the single dot charge coupling strength of the cavity containing the volume of the cavity $V$ and the material specific relative permittivity $\epsilon_r$ ($\epsilon_r = 16$ in Ge)~\cite{Cohen-Tannoudji1989, Burkard2006}.

\section{Spin rotations in the bonding state} \label{sec:spin_coupling}
The goal of this study is to define a qubit via the spin states in the low-energy bonding orbital. In Sec.~\ref{sec:bonding_anti_bonding} we showed that electric fields can couple bonding and anti-bonding states of equal spin and that the SOI perturbs the bonding spin states such that they couple to the anti-bonding state of opposite spin. It was argued that the combined effects of the cavity field and the SOI should allow for an effective spin coupling in the bonding orbital. In this section, we quantify this claim and show that in the logical qubit space defined by the perturbed bonding states, the total electric field-HH flopping mode qubit Hamiltonian may either be written in EDSR form,
	\begin{align}
	\label{eq:H_EDSR}
		H_{\text{EDSR}} = \frac{\Delta}{2} \sigma_z + \hbar \Omega_s \sin \left( \omega_d t \right) \sigma_y,
	\end{align}
when the qubit is driven by a classical field, or in Rabi form,
	\begin{align}
	\label{eq:qubit_Hamiltonian}
		H_{\text{cav}} = \frac{\Delta}{2} \sigma_z + \hbar \omega_c b^{\dagger} b + \hbar g_s \left(b +b^{\dagger}  \right) \sigma_y,
	\end{align}
for the case of interactions with single photons. The EDSR-qubit Hamiltonian~\eqref{eq:H_EDSR} may be used to perform single qubit operations, while two-qubit gates can be implemented with the cavity-qubit Hamiltonian~\eqref{eq:qubit_Hamiltonian}. As one- and two-qubit gates are universal for quantum computation~\cite{DiVincenzo1995}, the interactions described are sufficient to operate a complete quantum computer.

In the following we derive explicit expressions for the qubit energy separation $\Delta$ and the effective spin couplings $\Omega_s$ and $g_s$ for two special cases. Additionally, we highlight the strong dependence of the coupling on system parameters such as the inter-dot distance, the dot detuning and the applied magnetic field.

\subsection{Weak magnetic fields and spin-flip tunneling}
\label{subsec:weak_B}
Assuming inter-dot distances $2a \sim 100$~nm, we find $\vert \chi/ \eta \vert \ll 1$ for $B \leqslant 0.1$~T, allowing us to neglect $\chi$ at weak magnetic fields (cf. Fig.~\ref{fig:comparison_matrix_elements}). In this limit, we may exactly diagonalize the HH Hamiltonian in Eq.~\eqref{eq:H_GS_L_R} to obtain the eigenstates
	\begin{subequations}
	\begin{align}
		\label{eq:minus_up_bar}
		&\vert \overline{-, \Uparrow} \rangle = \cos \phi_- \vert -, \Uparrow \rangle - \sin \phi_-\vert +, \Downarrow \rangle, \\
		&\vert \overline{-, \Downarrow} \rangle = \cos \phi_+ \vert -, \Downarrow \rangle + \sin \phi_+  \vert +, \Uparrow \rangle, \\
		&\vert \overline{+, \Uparrow} \rangle = \cos \phi_+  \vert +, \Uparrow \rangle - \sin \phi_+  \vert -, \Downarrow \rangle, \\
		\label{eq:plus_down_bar}
		&\vert \overline{+, \Downarrow} \rangle = \cos\phi_- \vert +, \Downarrow \rangle + \sin \phi_- \vert -, \Uparrow \rangle,
	\end{align}
	\end{subequations}
where
    \begin{align}
        \phi_{\pm} = -\arctan  \frac{ \eta }{K_{\pm}+ \sqrt{K_{\pm}^2 + \eta^2}} 
    \end{align}        
are the spin-orbit mixing angles. 

\begin{figure}
		\includegraphics[scale=0.23]{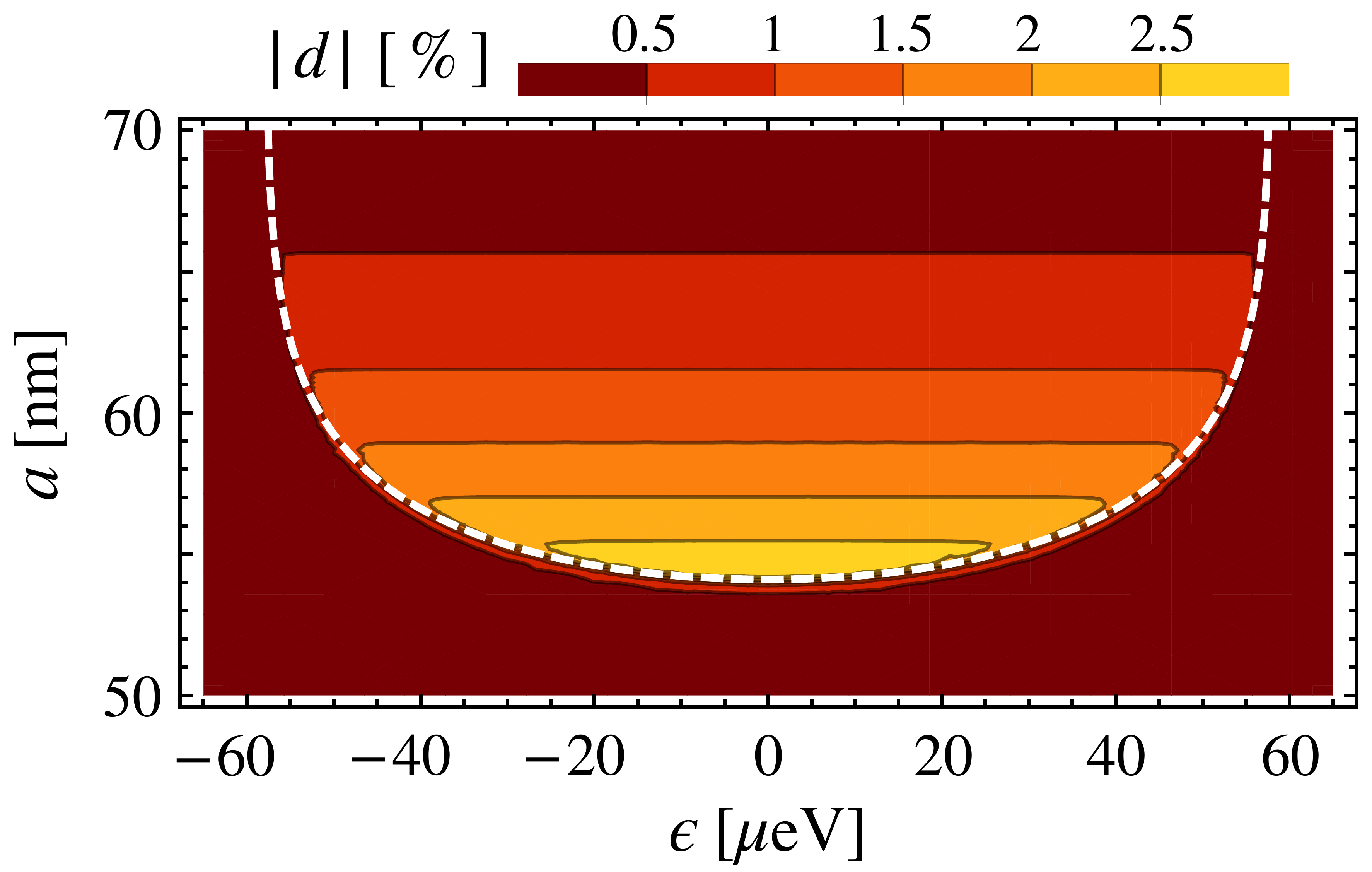}
		\caption{The relative spin coupling strength $\vert d \vert$ for both EDSR and photonic interactions as a function of the detuning $\epsilon$ and half the inter-dot distance $a$. There is a critical line where the coupling increases abruptly, described by the resonance condition $g \mu_B B = \sqrt{4t_c^2+ \epsilon^2}$ (white dashed line). We use $\hbar \omega_0 = 1$~meV, $g =10$, $\Xi = 100$~meV, $B=0.1$~T, $m=0.1 m_0$ and $\hbar \alpha_R \langle E_z \rangle = 10^{-11} $~eVm.}
		\label{fig:coupling_a_epsilon}
	\end{figure}

On the one hand, an electric field couples the bonding and anti-bonding states of equal spin, e.g., $\vert - , \Uparrow \rangle$ and $\vert + , \Uparrow \rangle$. On the other hand, the SOI couples bonding and anti-bonding states of opposite spin via spin-flip tunneling, e.g., $\vert - , \Uparrow \rangle$ and $\vert + , \Downarrow \rangle$ (cf. Eq.~\eqref{eq:minus_up_bar}). The combination of both effects therefore couples the logical states $\vert \overline{-, \Uparrow} \rangle$ and $\vert \overline{-, \Downarrow} \rangle$. This is seen most easily by expressing the interaction Hamiltonians in Eqs.~\eqref{eq:H_int_EDSR} and \eqref{eq:H_int_cavity} in the basis defined by Eqs.~\eqref{eq:minus_up_bar} -- \eqref{eq:plus_down_bar} in which the HH Hamiltonian $H$ is diagonal. At the level of the qubit space spanned by the logical basis $\lbrace \vert 0 \rangle = \vert \overline{-, \Downarrow} \rangle, \vert 1 \rangle = \vert \overline{-, \Uparrow} \rangle   \rbrace$, we find effective flopping-mode qubit Hamiltonians as given in Eqs.~\eqref{eq:H_EDSR} and \eqref{eq:qubit_Hamiltonian} with the effective spin couplings $\Omega_s = \Omega_c d$ and $g_s = g_c d$ containing the common dimensionless electric dipole matrix element
	\begin{align}
	\label{eq:momentum_operator_in_logical_SO_basis}
			d = d_c \sin \overline{\phi} + d_s  \left[ \frac{2\gamma}{1+\gamma^2} \cos \underline{\phi}  + \cos \overline{\phi}  \cos  \theta \right],
	\end{align}
where $d_c$ and $d_s$ are as defined in Eq.~\eqref{eq:dimensionless_momentum},  $\overline{\phi} = \phi_+ + \phi_-$ and $\underline{\phi} = \phi_+ - \phi_-$ (Appendix~\ref{appx:momentum_operator_in_spin_orbit_basis}). 
By definition, $d$ describes the ratio of spin to charge coupling strengths and its absolute value $ \vert d \vert$ is thus referred to as the relative spin coupling in what follows. It is shown as a function of the detuning $\epsilon$ and half the inter-dot distance $a$ in Fig.~\ref{fig:coupling_a_epsilon}. The coupling changes abruptly in a narrow region around the resonance $g \mu_B B = \sqrt{4t_c^2 + \epsilon^2}$, corresponding to the point where the energies of unperturbed bonding and anti-bonding states $\vert -, \Uparrow \rangle$ and $\vert +, \Downarrow \rangle$ align in energy. Beyond this critical line, i.e., when $E_{+ \Uparrow} < E_{- \Downarrow}$, the excited qubit state in Eq.~\eqref{eq:minus_up_bar} changes its character to be predominantly anti-bonding. This allows for strong dipole couplings to the qubit ground state, and the spin coupling mediated by the electric field is increased. Remarkably, the regime where the coupling becomes strongest is also the regime where the qubit is nearly ideally isolated from the rest of the Hilbert space, reducing leakage errors and hence defining an optimal operation mode (Appendix~\ref{appx:coupling_critical_line}).  We find the same form for the spin coupling if the tunnel matrix element is modulated via the in-plane confinement energy $\hbar \omega_0$ instead of the inter-dot distance (Appendix~\ref{appx:gs_function_of_omega_0}). The energies of the system read
	\begin{align}
		E_{\overline{\pm \sigma}} = \pm \sqrt{\eta^2 + \left(\pm \sqrt{t_c^2 + \epsilon^2/4} + \sigma g \mu_B B /2  \right)^2},
	\end{align}
where we define $\sigma = 1$ ($\sigma=-1$) for spin up (down) states, and the qubit energy gap is $\Delta = E_{- \Uparrow} - E_{- \Downarrow}$.  Hence, for $B = 0.1$~T we have $\Delta/ \hbar$ in the GHz range, allowing resonant coupling of the bonding ground state spin to photons confined in superconducting resonators~\cite{Burkard2020}.

\subsection{Symmetric dot configuration}
\label{subsec:symmetric_configuration}

As the most pronounced coupling strengths appeared at zero detuning, $\epsilon = 0$, in Sec.~\ref{subsec:weak_B}, we restrict our attention to the case of a symmetric DQD in this section, allowing us to consider arbitrary magnetic field strengths. Upon diagonalizing the HH Hamiltonian in Eq.~\eqref{eq:H_GS_L_R}, the new eigenenergies of the system read
	\begin{align}
 	E_{\overline{\pm \sigma}} = \pm \sqrt{(\sigma g \mu_B B/2 \pm  t_c)^2 + (\sigma \chi \mp  \eta )^2},
\end{align} 	
from which the  energy separation between the logical states can be easily computed as $\Delta = E_{\overline{- \Uparrow}} - E_{\overline{- \Downarrow}}$.
The spin states in the same orbital are degenerate at zero field, $B= 0$ due to $\chi (B=0) = 0$, in accordance with Kramers' theorem.  

\begin{figure}
		\includegraphics[scale=0.52]{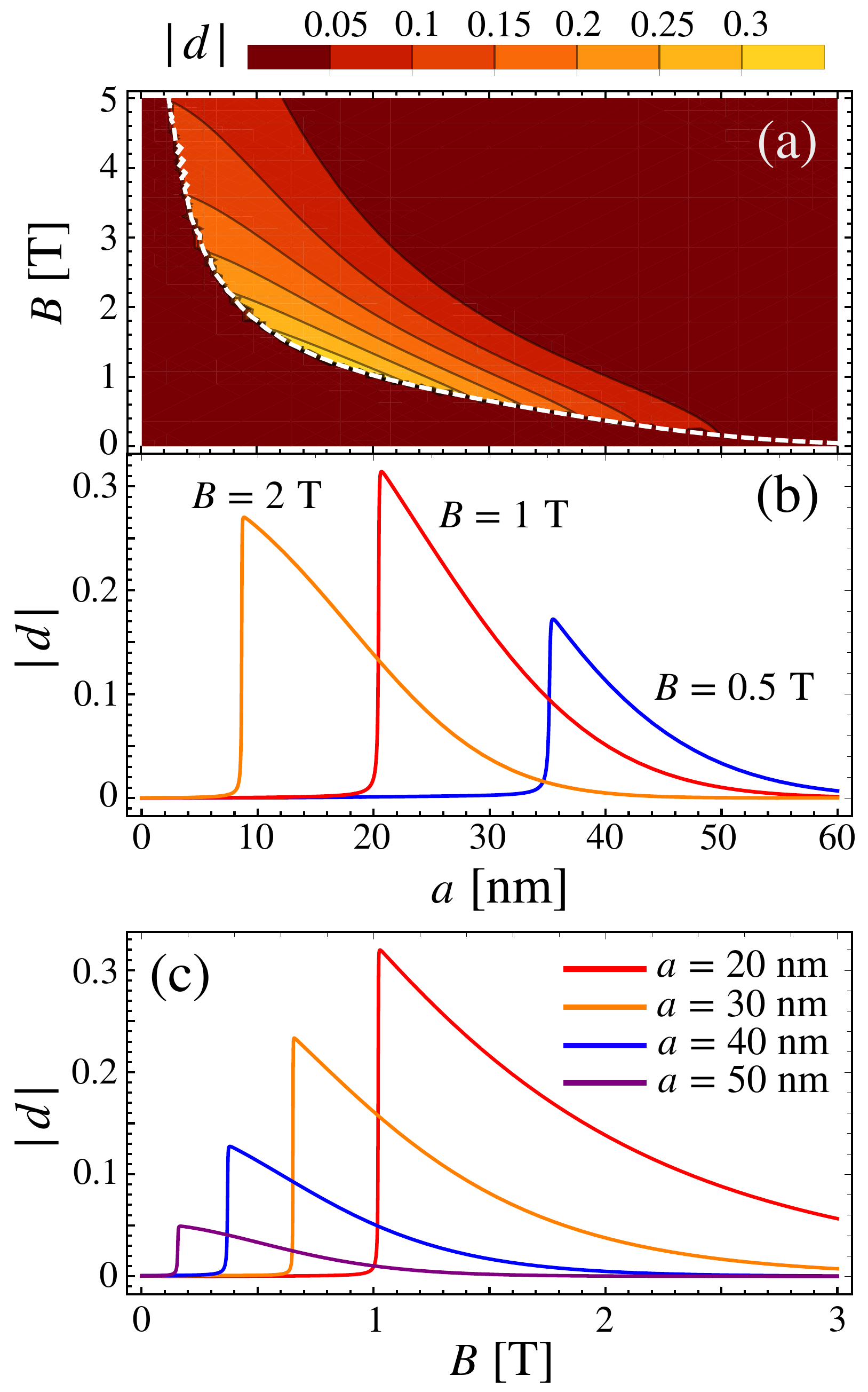}
		\caption{The relative spin coupling strength $\vert d \vert$ for both EDSR and photonic interactions. (a) Contour plot of $d$ as a function of half the inter-dot distance $a$ and the magnetic field $B$. The dashed white line shows the critical line $g \mu_B B = 2 t_c$ where the coupling changes abruptly. (b), (c) Line cuts at different magnetic fields (b) and dot separations (c) as indicated in the figure. Due to the magnetic compression of the oscillator states a smaller dot separation is favourable for larger magnetic fields. We find a characteristic step-like increase in the coupling when varying the control parameter, implying a high degree of control over the coupling strength. The remaining parameters are chosen as in Fig.~\ref{fig:coupling_a_epsilon}.}
		\label{fig:coupling_a_B}
	\end{figure}

Turning to the mixing between the qubit states, one may look at the eigenstates of \eqref{eq:H_GS_L_R}. They have the same form as in 
Eqs.~\eqref{eq:minus_up_bar} -- \eqref{eq:plus_down_bar} with the modified spin-orbit mixing angles
	\begin{align}
\phi_{\pm} \rightarrow \Phi_{\pm} =  -\arctan  \frac{\eta \mp  \chi}{K_{\pm} + \sqrt{K_{\pm}^2 + (\eta \mp  \chi)^2} } .
\end{align}
Consequently, we obtain the same form for the effective electric field mediated spin couplings $\Omega_s = \Omega_c d$ and $g_s = g_c d$ with $d$ as in Eq.~\eqref{eq:momentum_operator_in_logical_SO_basis} at $ \theta  = 0$. The relative coupling $\vert d \vert$ as a function of half the inter-dot distance and the applied magnetic field is displayed in Fig.~\ref{fig:coupling_a_B}. As can be seen from the plot, the coupling strength shows a strong dependence on the applied magnetic field and the inter-dot distance. Indeed, there exists a critical line where the coupling changes abruptly. It is described by the resonance condition $g \mu_B B = 2t_c$, which is explained physically as in Sec.~\ref{subsec:weak_B} by equal energies of unperturbed bonding and anti-bonding states and a change of character in the excited qubit state when the line is crossed (Appendix~\ref{appx:coupling_critical_line}). The subsequent decrease in $\vert d \vert$ is due to the exponential decay of the dipole elements $d_c$ and $d_s$ in both $a$ and $B$ (Eq.~\eqref{eq:dimensionless_momentum}). This knowledge can be taken into account in QD manufacturing to maximize the relative spin coupling strength $\vert d \vert$ by constructing the system such that the inter-dot distance equals $2a_{\text{max}}$, where $a_{\text{max}}$ is the value of $a$ where $\vert d \vert$ becomes maximum for a fixed magnetic field. In a DQD where the inter-dot distance is tunable (or, alternatively, the in-plane confinement energy, see Appendix~\ref{appx:gs_function_of_omega_0}), the characteristic step form of $\vert d \vert$ may be utilized as a spin-orbit switch by only slightly changing $a$ (Fig.~\figref[(b)]{coupling_a_B}). Since the dependence of $d$ on the magnetic field shows a similar behaviour (Fig.~\figref[(c)]{coupling_a_B}), varying the magnetic field may also be used to control the qubit-field interaction. Such tunable couplings are of great importance in quantum processing units. Large couplings can be used to quickly manipulate the spin qubit state, while low couplings may be used to isolate a given spin state, e.g., for the purpose of  qubit read-out or initialization.
\subsection{Feasible qubit gate times}
\label{subsec:feasible_gate_times}
Finally, we may estimate one- and two-qubit gate times which may be achieved with a natural HH flopping mode qubit in present-day experiments. We assume the system to be engineered such that the qubit is operated at the optimal point, where we may reach relative spin couplings $\vert d \vert > 1/4$ (cf. Fig.~\ref{fig:coupling_a_B}).

One-qubit gates are most efficiently driven by an alternating electric field of magnitude $\mathcal{E}$ and frequency $\omega_d$. For a QD of confinement energy $\hbar \omega_0 = 1$~meV (yielding a lateral size of $r \sim 25$~nm) and an electric field at the resonant driving frequency $\omega_d = \Delta \sim 100$~GHz at $B = 1$~T, we find $\Omega_s \sim (1- 10)$~$\mathcal{E}$~[V/m]~MHz and thus one-qubit gate times $\tau_1 \sim 1/\Omega_s $ of less than a nanosecond for $\mathcal{E} \sim (0.1 - 1)$~kV/m.

On the other hand, two-qubit gates may be implemented by harnessing the long-range spin-photon coupling introduced by a bosonic cavity to obtain a controlled interaction between distant spins. For instance, the iSWAP gate may be performed in the dispersive regime in time $\tau = (4k+1)\pi \vert \delta \vert /2g_s^2$ with spin qubit-cavity detuning $\delta = \omega_c - \Delta/\hbar$ and $k = 0,1,2,\dots$~\cite{Benito2019a}. The entangling and universal CNOT gate may then be implemented using two iSWAP gates and one-qubit gates~\cite{Barenco1995, Schuch2003}. As superconducting resonators have typical coupling strengths of $ g_c \omega_c/ 2\pi \omega_0 \simeq 1-10$~MHz for QDs of lateral size $r \simeq 10-100$~nm~\cite{Burkard2020}, HH flopping mode systems in Ge allow for fast two-qubit logic with typical gate operation times $\tau_2 \sim \vert \delta \vert $~[kHz]~ns.

\section{Conclusion} 
\label{sec:conclusion}

We show that a flopping mode qubit arises naturally for HHs in Ge, i.e., without the need to create synthetic SOI via a magnetic field gradient. This is made possible by the strong cubic Rashba SOI, which leads to intra-dot spin flips and inter-dot spin-flip tunneling. Utilizing these processes, we derive an effective electric field mediated spin coupling of the bonding state for different parameter regimes. Additionally, we quantify and physically motivate optimal points of operation and argue that the spin coupling is highly tunable and can reach several ten percent of the charge coupling when the system is engineered accordingly. Consequently, both one- and two-qubit gate times in the nanosecond range are feasible in present-day experiments with natural HH flopping mode qubits. Our results highlight yet another possible application in quantum computing of the versatile platform Ge.

Further research may include the generalization of the system geometry to elliptical QDs where the effect of the SOI is expected to become more diverse due to a less restrictive symmetry, i.e., the SOI may couple the ground state of a given spin to more than one excited level. Moreover, typical relaxation times for flopping mode systems realized with HHs in Ge need to be calculated to estimate their performance under realistic conditions and definitively judge their potential as fast and reliable qubits.

\onecolumngrid

\appendix

\section{Fock-Darwin states}
\label{appx:Fock-Darwin_states}

The Fock-Darwin states $\psi^{\text{FD}}_{nl}(x,y)$ are eigenstates of the Hamiltonian
    \begin{align}
        H_{\text{FD}} = \frac{\pi_x^2+ \pi_y^2}{2m} + \frac{1}{2}m \omega_0 (x^2+y^2),
    \end{align}
where $\boldsymbol{\pi} = \mathbf{p} + e  \mathbf{A}$ with $\mathbf{A}=B(-y,x,0)/2$ in the symmetric gauge. Due to the circular symmetry of the system, it is convenient to express the general wave functions in planar polar coordinates $(r, \varphi)$,
	\begin{align}
		\psi_{nl}(r, \varphi) = \sqrt{\frac{n!}{\pi(n+ \vert l \vert )!}} \frac{e^{il \varphi}}{b} \left( \frac{r}{b} \right)^{\vert l \vert} L_n^{\vert l \vert} \left( \frac{r^2}{b^2}  \right) e^{-r^2/2b^2},
	\end{align}
where we introduce the main and azimuthal quantum numbers $n$ and $l \in \lbrace -n, - n+2, ..., n-2,n \rbrace$, respectively, $L_n^{\vert l \vert} \left( r^2/b^2  \right)  = (-1)^{\vert l \vert} \partial_r^{\vert l \vert} L_{n + \vert l \vert}\left( r^2/2b^2  \right)$ denote the generalized Laguerre polynomials and $b^2 = \hbar / m  \sqrt{\omega_0^2+\omega_L^2}$ with Larmor frequency  $\omega_L = eB/2m$ \cite{Fock1928,Darwin1930}. In the main text, we only need the ground state ($n=0$) and the third excited states with extremal azimuthal quantum number ($n=3, l = \pm 3)$. They read in cartesian cooridnates $(x,y)$,
    \begin{align}
        &\psi^{\text{FD}}_{00} = \frac{1}{\sqrt{\pi}b}e^{-(x^2+y^2)/2b^2}, \\
        &\psi^{\text{FD}}_{3 \pm3} = \frac{1}{\sqrt{6\pi}b^4 } (x \pm iy)^3  e^{-(x^2+y^2)/2b^2}.
    \end{align}
\section{Matrix elements in the spin-flip tunneling term}
\label{appx:spin_flip_matrix_elements}

In this appendix we display the matrix elements needed for the calculation of the correction to the spin-flip tunneling term $\delta \tilde{\eta}_1$ in Eq.~\eqref{eq:matrix_elements_excited_states} of the main text. Writing $h_0 = H_0 - g \mu_B B \sigma_z/2$, they read
	\begin{align}
	\begin{split}
		& \langle \psi_{00}^L \vert h_0 \vert \psi_{3 \pm 3}^R\rangle= \int_{\mathbb{R}^2} \text{d}x \; \text{d}y \;  \psi_{00}^{L*} h_0  \psi_{3 \pm 3}^R = S\frac{\sqrt{3 \hbar/2 m \omega}}{32 a \hbar \omega^3} \Bigg\lbrace \frac{4 a^4m^2 \omega_0^6}{\hbar} \left( 5 \hbar +a^2m \left(\omega \mp 3 \omega_L \right) \right) \\
		&  + \hbar^2 \omega_0^4 \bigg[ -6 + \frac{4a^4 m^2 \omega_L}{\hbar^2}(17 \omega_L \mp 11 \omega) + \frac{4a^6m^3 \omega_L^2}{h^3} (5 \omega \mp 7 \omega_L) + \frac{a^2 m}{\hbar} (11 \omega \mp 9 \omega_L)\bigg]\\
		&  + 2\hbar \omega_0^2 \omega_L \bigg[ \mp\frac{ 8 a^6 m^3\omega_L^4}{\hbar^2} \pm 3\hbar \omega + \frac{8 a^4 m^2 \omega_L^3}{\hbar} \left(3+ \frac{a^2 m \omega}{\hbar} \right)  \\
		& +\hbar \omega_L \left(\mp 3 + \frac{4a^2 m \omega}{\hbar} \right) \mp 4a^2 m \omega_L^2 \left( 1 + \frac{6a^2 m \omega}{\hbar} \right) \bigg] \Bigg\rbrace -S a^3 \sqrt{6} \left( \frac{m}{\hbar \omega} \right)^{3/2} \omega_{\mp}^3 \left( 4 \hbar \omega \pm 3 \hbar \omega_L \right).
	\end{split}
	\end{align}

\section{Wannier states including excited orbitals}

\subsection{Numerical computation of matrix elements}
\label{appx:numerical_study}

\begin{figure*}
		\includegraphics[scale=0.27]{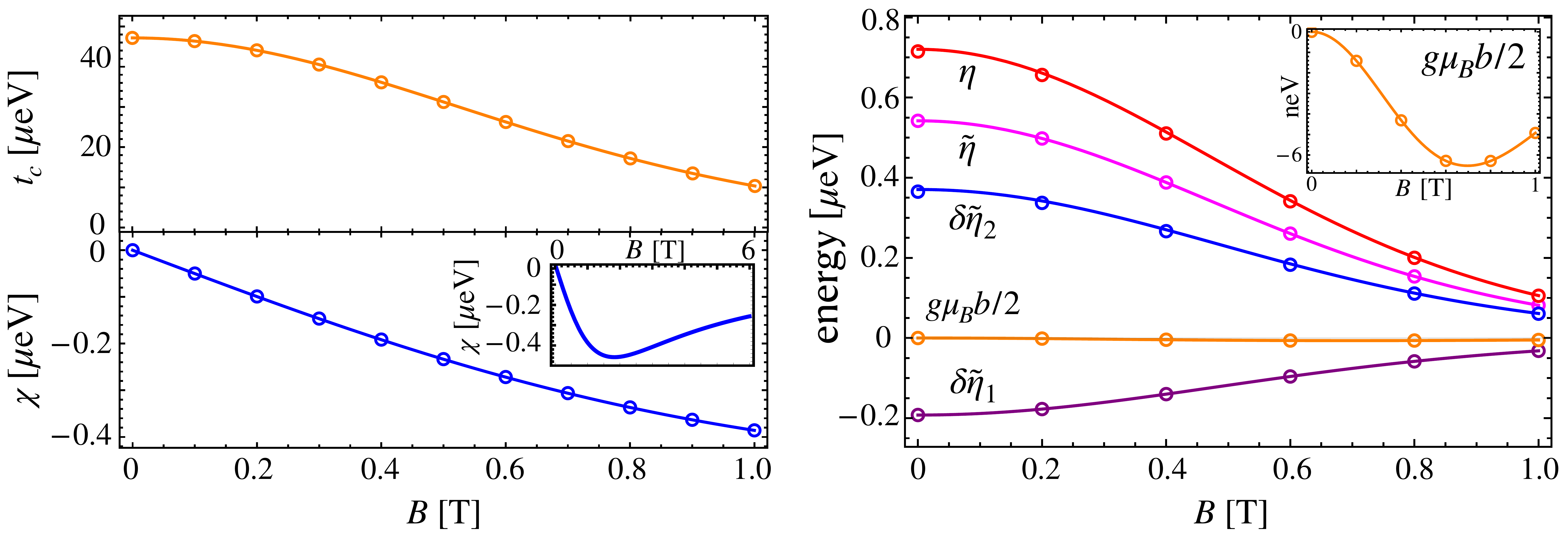}
		\caption{Comparison of the matrix elements in the DQD system. We show the tunneling element $t_c$ and the spin-orbit induced spin-flip terms $\chi$ and $ \eta$ as given in Eqs.~\eqref{eq:matrix_elements} and \eqref{eq:matrix_elements_excited_states} as a function of the magnetic field $B$ (solid lines) and the results of the numerical study (circles). For the spin-flip tunneling term $\eta$ we display the constituent terms separately to highlight the high accuracy of the approximative analytical results. We use the system parameters $\hbar \omega_0 = 1$~meV, $a =50$~nm, $\Xi = 100$~meV, $m=0.1m_0$ and $ \hbar \alpha_R \langle E_z \rangle = 10^{-11} $~eVm.}
		\label{fig:comparison_matrix_elements}
	\end{figure*}

In our numerical study we incorporate the overlaps between states of different pseudo-spin, resulting in the overlap matrix $M_S$ in the basis $\left\lbrace{ \vert \tilde{\psi}_{\Uparrow}^L \rangle},  { \vert \tilde{\psi}_{\Downarrow}^L \rangle},  
{ \vert \tilde{\psi}_{\Uparrow}^R \rangle},  
{ \vert \tilde{\psi}_{\Downarrow}^R \rangle} \right\rbrace$ defined in Sec.~\ref{sec:effect_higher_orbitals} of the main text,
	\begin{align}
		M_S = \mathds{I} + \delta M_S =  \begin{pmatrix}
			1 & 0 & S_{\Uparrow} & S_{\Uparrow \Downarrow} \\
			0 & 1 & -  S_{\Uparrow \Downarrow} & S_{\Downarrow}  \\
			S_{\Uparrow} & -S_{\Uparrow \Downarrow} & 1 & 0 \\
			 S_{\Uparrow \Downarrow} & S_{\Downarrow} & 0 & 1
		\end{pmatrix},
	\end{align}
where the off-diagonal elements collected in $\delta M_S$ are much smaller than unity in the present system. Aiming to orthogonalize the states, we introduce the Wannier matrix $W$ connecting the left and right dot eigenstates to the Wannier states,
	\begin{align}
		\begin{pmatrix}
		\vert L, \Uparrow \rangle \\
		\vert L, \Downarrow \rangle \\
		\vert R, \Uparrow \rangle \\
		\vert R, \Downarrow \rangle 
		\end{pmatrix}		
		= W \begin{pmatrix}
	 { \vert \tilde{\psi}_{\Uparrow}^L \rangle} \\
		 { \vert \tilde{\psi}_{\Downarrow}^L \rangle} \\
		 { \vert \tilde{\psi}_{\Uparrow}^R \rangle} \\
		 { \vert \tilde{\psi}_{\Downarrow}^R \rangle}
		\end{pmatrix}		
	\end{align}
The orthonormalization condition $\langle d, \sigma \vert  d', \sigma' \rangle = \delta_{dd'} \delta_{\sigma \sigma'}$ for $d \in \lbrace L,R \rbrace$ and $\sigma \in \lbrace \Uparrow,\Downarrow \rbrace$ can be moulded into the Wannier equation in matrix form,
	\begin{align}
	\label{eq:Wannier_eq}
		WM_SW^{\dagger} = \mathds{I}.
	\end{align}
Solving this system of non-linear equations yields the components of $W$ and thus the Wannier states. We choose the physical solution given by the one with the largest diagonal elements. With the Wannier states at hand, one may then compute the matrix elements by numerical integration for different values of the magnetic field. A comparison between the numerical approach and the analytical results given in Eqs.~\eqref{eq:matrix_elements} and \eqref{eq:matrix_elements_excited_states} is shown in Fig.~\ref{fig:comparison_matrix_elements}.

\subsection{Approximative solution to the Wannier equation}
\label{appx:approximative_solution_Wannier}

\begin{figure}
		\includegraphics[scale=0.55]{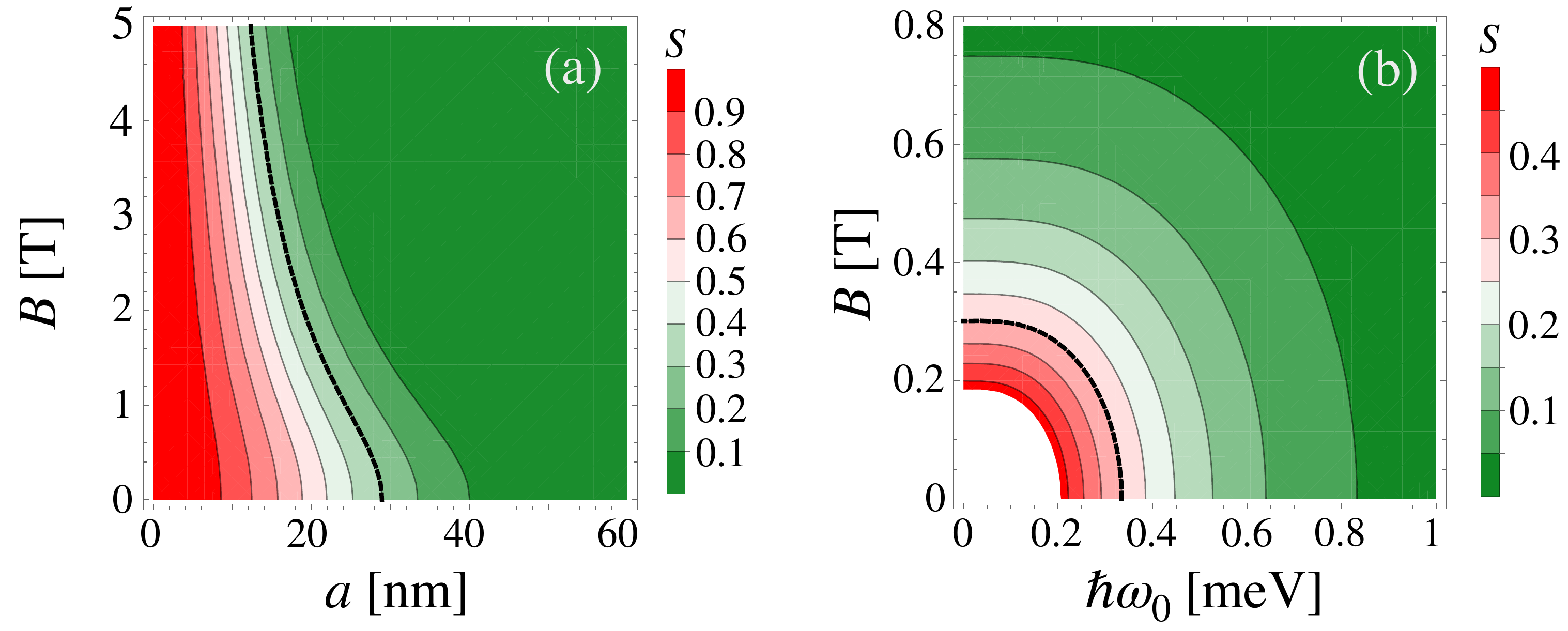}
		\caption{Validity of the approximative solution to the Wannier equation. We show the overlap $S$ as a function of the magnetic field $B$ and (a) half the dot separation $a$ and (b) the in-plane confinement energy $\hbar \omega_0$. The approximation is valid in the green region, while it does not accurately describe the system in the red region. For $S \lesssim 0.3$ (to the right of the dashed line) the error caused by neglecting second-order terms is less than ten percent. We use the system parameters $g =10$, $\Xi = 100$~meV, $m=0.1m_0$ and $\hbar \alpha_R \langle E_z \rangle = 10^{-11} $~eVm as well as (a) $\hbar \omega_0 = 1$~meV and (b) $a=50$~nm.}
		\label{fig:validity_of_approximation}
	\end{figure}

\begin{figure*}
		\includegraphics[scale=0.26]{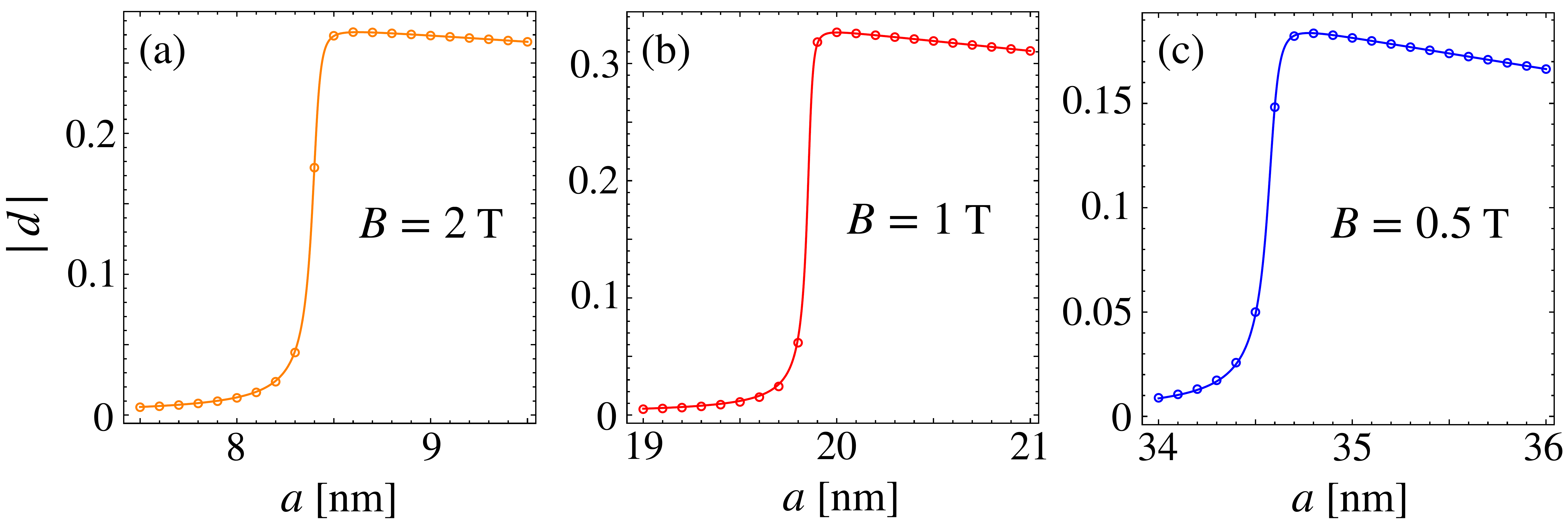}
		\caption{Comparison between approximated analytical (solid lines) and exact numerical results (circles) for the relative spin coupling strength $\vert d \vert$ as a function of the magnetic field $B$ as given by the line cuts in Fig.~\figref[(b)]{coupling_a_B}. In each plot we display the region of half the inter-dot distance $a$ in which $\vert d \vert$ shows the characteristic step form. We find excellent agreement between analytics and numerics even for $a<10$~nm. The system parameters are set to $\hbar \omega_0 = 1$~meV, $g =10$, $\Xi = 100$~meV, $m=0.1m_0$ and $\hbar \alpha_R \langle E_z \rangle = 10^{-11} $~eVm.}
		\label{fig:coupling_vs_a_numerical_check}
	\end{figure*}

When comparing the numerical data to the approximative solutions of Sec.~\ref{sec:effect_higher_orbitals}, we find that all terms are in excellent agreement except for a correction to the spin-flip tunneling element due to a non-vanishing overlap between states of different pseudo-spin,
	\begin{align}
	\label{eq:spin_flip_Wannier_overlap}
		\delta \tilde{\eta}_2 = W_{22}W_{32} \left( \langle \psi_{00}^L \vert h_0 \vert \psi_{00}^L \rangle + \langle \psi_{00}^R \vert h_0 \vert \psi_{00}^R \rangle \right),
	\end{align}
where $h_0 = H_0 - g \mu_B B/2$ and $W_{nm}$ denote the components of the Wannier matrix $W$. We may approximate the solution to Eq.~\eqref{eq:Wannier_eq} as $W  = \mathds{I} - \delta M_S/2$, resulting in an error only at quadratic order in the small off-diagonal elements, $W M_S W^{\dagger}  = \mathds{I} + \mathcal{O}(\delta M_S^2)$. Consequently, the spin-flip correction reads
	\begin{align}
	\label{eq:spin_flip_Wannier_overlap_2}
		\delta \tilde{\eta}_2 =  \frac{6 \lambda_R S  m^3a^3 \omega_0^6}{ \omega^2 \left( 9\omega_0^2 +6 \omega_L \omega_Z - \omega_Z^2 \right)} \left( \omega + \frac{3 \hbar \omega_0^2}{32 a^2 m\omega^2}  \right),
	\end{align}
which is the form displayed in Eq.~\eqref{eq:matrix_elements_excited_states} of the main text. The approximation is expected to hold for small overlaps between different dot states, and we naively expect good agreement with the exact numerical results for $a  \gtrsim 30 $~nm (Fig.~\ref{fig:validity_of_approximation}). Note, however, that this bound is too pessimistic as we can solve the Wannier equation exactly for $ S_{\Uparrow \Downarrow} = 0$ and only have to resort to the approximative solution for the computation of $\delta \tilde{\eta}_2$, which tends to zero as the dot separation is decreased. As a consequence, even for dot separations smaller than $2a = 60$~nm the spin coupling is not changed noticeably as can be seen from Fig.~\ref{fig:coupling_vs_a_numerical_check}.

\section{Bonding-anti-bonding basis}
\label{appx:bonding_anti_bonding_basis}

The bonding-anti-bonding basis is defined to be the eigenbasis of the Hamiltonian $H = -\frac{\epsilon}{2} \tau_z - t_c \tau_x$, where the operators $\tau$ are Pauli matrices with respect to the left and right dot states with the convention $\tau_z = \vert L \rangle \langle L \vert - \vert R \rangle \langle R \vert$. One finds
	\begin{align}
		U H U^{\dagger} = \text{diag} \left(-E,  E \right),
	\end{align}
where $E = \sqrt{t_c^2+ \epsilon^2/4}$ and
	\begin{align}
		U = \begin{pmatrix}
			\cos \vartheta &   \sin \vartheta  \\
			- \sin \vartheta &  \cos \vartheta  \\
		\end{pmatrix},
	\end{align}
with the hybridization angle $\vartheta =  \arctan \left( 2t_c/\left(\epsilon + \sqrt{\epsilon^2+4t_c^2} \right) \right)$. Since $\vartheta \in [0, \pi/2]$ for $t_c >0$, it may be rewritten as $\vartheta  = \pi/4 - \theta/2$ with the orbital angle $\theta = \arctan( \epsilon/2t_c)$.

\section{Momentum operator in spin-orbit basis}
\label{appx:momentum_operator_in_spin_orbit_basis}

The dimensionless momentum operator in the basis defined by Eqs.~\eqref{eq:minus_up_bar}--\eqref{eq:plus_down_bar} in Sec.~\ref{subsec:weak_B} of the main text has the form
	\begin{align}
	\label{eq:momentum_operator_in_SO_basis}
	\begin{split}
		\Pi_x = \left( d_c \sin \overline{\phi}  + d_s \cos \overline{\phi}  \cos  \theta \right) \tau_z \sigma_y + \frac{2\gamma}{1+\gamma^2} d_s \left(\cos \underline{\phi} \sigma_y + \sin \underline{\phi}  \tau_y \sigma_z\right) + d_s \sin  \theta \tau_x \sigma_y  + \left( d_c \cos \overline{\phi} - d_s \sin \overline{\phi}  \cos  \theta \right) \tau_y,
	\end{split}
	\end{align}
where $\overline{\phi} = \phi_+ + \phi_-$ and $\underline{\phi} = \phi_+ - \phi_-$.  In the logical space defined by the pseudospin in the bonding state, one finds $\Pi_x = d \sigma_y$ with
	\begin{align}
			d = d_c \sin \overline{\phi} + d_s  \left[ \frac{2\gamma}{1+\gamma^2} \cos \underline{\phi}  + \cos \overline{\phi}  \cos  \theta \right]. 
	\end{align}	 
For the symmetric dot configurations discussed in Sec.~\ref{subsec:symmetric_configuration} the form is the same after setting $ \theta  = 0$ and substituting $\phi_{\pm} \rightarrow \Phi_{\pm}$.

\section{Critical line of the spin coupling strength}
\label{appx:coupling_critical_line}

The spin-orbit mixing angles are of the form $\nu_{\pm} =   \arctan \Lambda_{\pm}$ with
	\begin{align}
	\label{eq:Lambda}
		\Lambda_{\pm} = \frac{f_{\pm}}{h \pm g \mu_B B/2 +  \sqrt{f_{\pm}^2 +(h \pm g \mu_B B/2 )^2 }},
	\end{align}
where $f_{\pm} = \eta$ and $h = \sqrt{t_c^2+ \epsilon^2/4}$  ($f_{\pm} = \eta \mp \chi$ and $h=t_c$) in Sec.~\ref{subsec:weak_B} (Sec.~\ref{subsec:symmetric_configuration}). Since $d_s \ll d_c$, the relative spin coupling strength is of the form
	\begin{align}
		d \simeq  d_c \sin (\nu_- + \nu_+) \simeq d_c \sin \nu_-,
	\end{align}
where the last approximation is valid due to the larger denominator in $\nu_+$ for $B>0$ and $f_{\pm} \ll h$. For $g \mu_B B \lessgtr 2h$ one has $ \Lambda_- \lessgtr 1$. Moreover, since $ \Lambda_- > 0$, the corresponding spin-orbit mixing angle may be written as
    \begin{align}
        \nu_-^< =   \frac{1}{2} \arctan \lambda_-, \; \nu_-^> =   \frac{1}{2} \arctan \lambda_- \pm \frac{\pi}{2}, 
    \end{align}
where the sign in $\nu_-^>$ is positive (negative) for $f_- >0$ ($f_- < 0)$ and   
    \begin{align}
         \lambda_- =  \frac{f_{\pm}}{h \pm g \mu_B B/2}.
    \end{align}
Since $\lambda_- \ll 1$ outside a narrow region around the resonance $g \mu_B B = 2h$, $\vert \sin \xi_- \vert $ displays a characteristic step-like behaviour. Consequently, the relative spin coupling also shows this behaviour, accompanied by the exponential decay of $d_c$ in both $a$ and $B$. The critical line $g \mu_B B = 2h$, i.e., the point where the coupling strength changes abruptly is shown as dashed white lines in the contour plots in Secs.~\ref{subsec:weak_B} and \ref{subsec:symmetric_configuration}.

The physical reason for the critical line is that the unperturbed states $\vert -, \Uparrow \rangle $ and $\vert +, \Downarrow \rangle $ align in energy. When the energy of the state $\vert +, \Downarrow \rangle$ is reduced further, the excited qubit state (Eq.~\eqref{eq:minus_up_bar}) changes its character as the anti-bonding state $\vert +, \Downarrow \rangle$ becomes the dominant contribution. Hence, it becomes more and more susceptible to the electric dipole coupling with the state $\vert -, \Downarrow \rangle$ which is the dominant contribution to the ground qubit state $\vert \overline{-, \Downarrow} \rangle$.

We remark that beyond the critical line, the qubit space is almost ideally isolated from the rest of the Hilbert space under consideration. Using that $d_s$ is negligibly small compared to $d_c$ and taking into account the above considerations, the momentum operator in the spin-orbit basis reads (Appendix~\ref{appx:momentum_operator_in_spin_orbit_basis}),
    \begin{align}
        \Pi_x \simeq d_c \sin \nu_- \tau_z \sigma_y + d_c \cos \nu_- \tau_y.
    \end{align}
Since $\nu_- \simeq \pm \pi/2$ beyond the critical line, one has $\cos \nu_- \simeq 0$, and the qubit space is decoupled in good approximation.

\section{Spin coupling with variable confinement energy}
\label{appx:gs_function_of_omega_0}

\begin{figure}
		\includegraphics[scale=0.28]{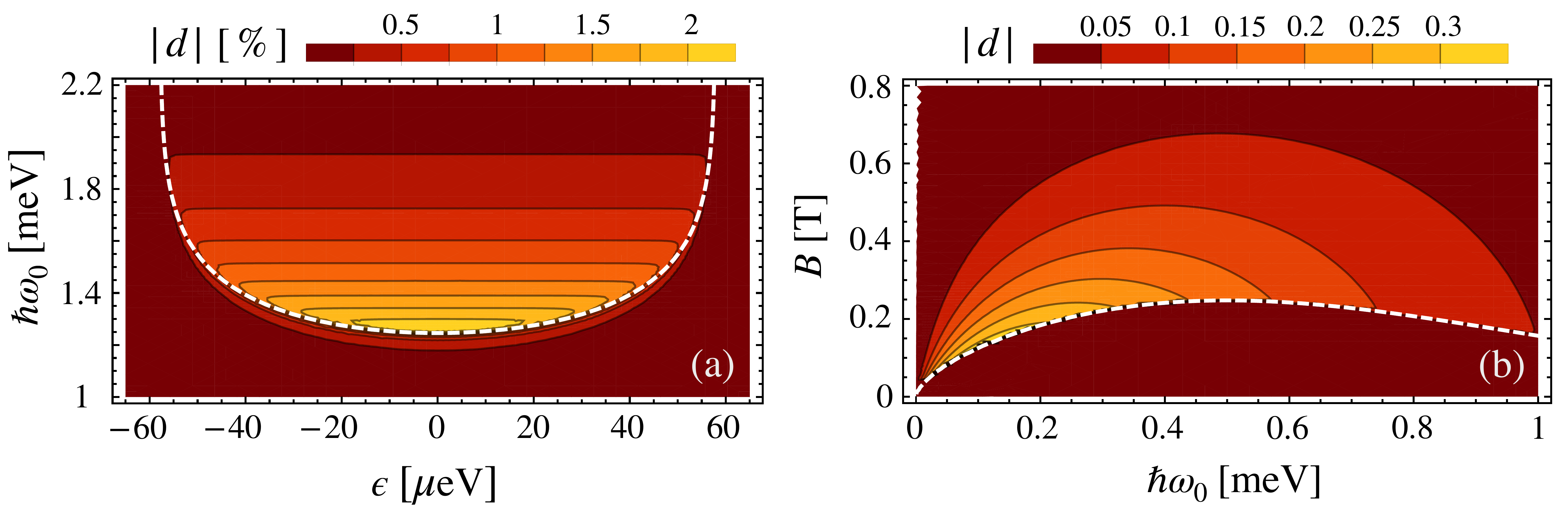}
		\caption{Relative pin coupling strength $\vert d \vert$ as a function of the in-plane confinement energy $\hbar \omega_0$ and (a) the detuning $\epsilon$ (b) the applied magnetic field $B$ corresponding to the special cases discussed in Sec.~\ref{subsec:weak_B} and Sec.~\ref{subsec:symmetric_configuration}, respectively. The dashed white lines indicate the critical lines and are described by the equations (a) $\sqrt{4t_c^2 + \epsilon^2} = g \mu_B B$ and (b) $2t_c = g \mu_B B$ as detailed in Appendix~\ref{appx:coupling_critical_line}. We set $g =10$, $\Xi = 10$~meV, $m=0.1m_0$ and $\alpha_R \langle E_z \rangle = 10^{-12} $~eVm and $a=50$~nm.}
		\label{fig:coupling_w0}
	\end{figure}

In this appendix we consider the spin coupling strengths derived in Secs.~\ref{subsec:weak_B} and \ref{subsec:symmetric_configuration} but with a fixed dot separation. Instead, the tunability of the tunnel coupling $t_c$ is taken into account by varying the in-plane confinement energy $\hbar \omega_0$. The results are shown in Fig.~\ref{fig:coupling_w0}.
	
\twocolumngrid

\bibliographystyle{apsrev4-2}

\bibliography{Ge_holes_literature}

\end{document}